\documentclass[a4paper]{jpconf}
\usepackage[utf8]{inputenc}

\usepackage{lmodern}
\usepackage{graphicx}
\usepackage{bm}
\usepackage{hyperref}
\hypersetup{linktocpage,colorlinks,citecolor={blue},pdfdisplaydoctitle=true,pdfpagemode=UseOutlines,bookmarksnumbered=true}
\usepackage{mathrsfs,dsfont}
\usepackage{color}
\usepackage{amsmath}
\usepackage{cite}
\usepackage{doi}

\newcommand{\bra}[1]{\langle #1 |} 
\newcommand{\ket}[1]{| #1 \rangle } 
\newcommand{\upd}{\mathrm{d}}
\newcommand{\ie}[0]{\textit{i.e.} }
\newcommand{\eg}[0]{\textit{e.g.} }
\newcommand{\xb}[0]{\mathbf{x}}
\newcommand{\yb}[0]{\mathbf{y}}

\newcommand{\rhoav}[0]{\bar{\rho}}

\begin{document}
\title{Does gravity have to be quantized? Lessons from non-relativistic toy models}
\author{Antoine Tilloy}
\address{Max-Planck-Institut f\"ur Quantenoptik, Hans-Kopfermann-Stra{\ss}e 1, 85748 Garching, Germany}
\ead{antoine.tilloy@mpq.mpg.de}

\begin{abstract}
It is often argued that gravity has to be a quantum theory simply because a fundamentally semiclassical approach would necessarily be inconsistent. Here I review recent Newtonian toy models of (stochastic) semiclassical gravity. They provide one option to implement a force semiclassically without getting into the known problems associated with mean-field. These models are not complete theories and should not be considered too seriously, but their consistency shows that semiclassical gravity is hard to dismiss on purely theoretical grounds.
\end{abstract}

\section{Introduction}

Do we have to quantize gravity? Is it merely possible that gravity is not quantum? Are there theoretical arguments or experimental results cornering us to the quantization route? Is it already known that gravity must be a quantum theory and that only its specific form is yet to be understood? 

My point, which is not particularly original, is simply that we are not sure. The situation is not as clear cut as one might think and the arguments in favor of the quantization of gravity are so far weak. Naturally, I do not aim to demonstrate that gravity is not quantum with certainty, but merely that this is a possibility one ought to consider. I am obviously not the first one to defend this line, and the possibility that gravity is ultimately not quantum has abundantly been discussed in the literature \cite{carlip2001,mattingly2005,wuthrich2005}. My objective here is simply to review recent constructions of consistent Newtonian toy models \cite{kafri2014,kafri2015, tilloy2016,tilloy2017,tilloy2017grw,tilloy2018_binding,khosla2017,altamirano2018noclassicalchannel} that exemplify how standard arguments against semiclassical gravity can be bypassed. 

The present contribution to these proceedings does not contain anything fundamentally new on the mathematical side, but puts together all the arguments and derivations, in an hopefully selfcontained way. It follows to a large extent, although in a more detailed manner, the rather informal discussion in \cite{tilloy2018_binding}.

\subsection{Semiclassical and fully quantum gravity}
So what is the alternative to a fully quantum theory of gravity? Because the words \emph{classical}, \emph{semiclassical}, and \emph{quantum} have been used to mean different things in this context, a clarification is perhaps not superfluous. In a classical theory of gravity, one has a spacetime, classical (well localized) and its properties are encoded in a metric $g(t,\xb)$. Matter is classical, one has particles with well defined trajectories or \eg a scalar field $\phi(t,\xb)$. The two evolve jointly: the dynamics of matter is influenced by the metric (essentially through the substitution $\partial_\mu \rightarrow D_\mu$), and matter in turn curves space time via Einstein's equation $G_{\mu\nu}\propto T_{\mu\nu}$. In a fully quantum theory of gravity, say canonical quantum gravity to fix the ideas, both sides become quantum and a wave function $\psi_t(\phi,g)$ is needed to describe the couple matter + spacetime.

In a semiclassical theory of gravity, spacetime is still classical, with a well defined metric $g(\xb,t)$. Matter is quantum, and is described at an intuitive and naive level by some wave-function $\psi_t(\phi)$. Its dynamics is given by Quantum Field Theory in curved spacetime (QFTCST), which is naively given by the transformation $\partial_\mu \rightarrow D_\mu$ in Lorentzian QFT, and is a refined version of ordinary quantum mechanics in an external classical field. The difficult part is then to find a law for the backreaction of quantum matter on a classical spacetime: when there are superpositions of matter, how is spacetime curved? While QFTCST for a fixed metric $g(\xb,t)$ is uncontroversial and even experimentally well tested in the non-relativistic limit (e.g. by letting quantum stuff fall), the backreaction problem is really where wild speculations are allowed. 
A particular proposal, sometimes called semiclassical Einstein equations, M{\o}ller-Rosenfeld semiclassical gravity \cite{moller1962,rosenfeld1963}, or sometimes simply semiclassical gravity, is to have spacetime be curved by the quantum expectation values, i.e. $G_{\mu\nu}\propto \langle \hat{T}_{\mu\nu}\rangle$. Heuristically, at the non-relativistic level, it means having the modulus squared of the wave-function source the gravitational field: if the wavefunction for a single particle has two blobs, they attract each other. Note again that this is just \emph{one} option to curve a classical spacetime with quantum matter. 

\begin{figure}
    \centering
    \includegraphics[width=0.32\textwidth]{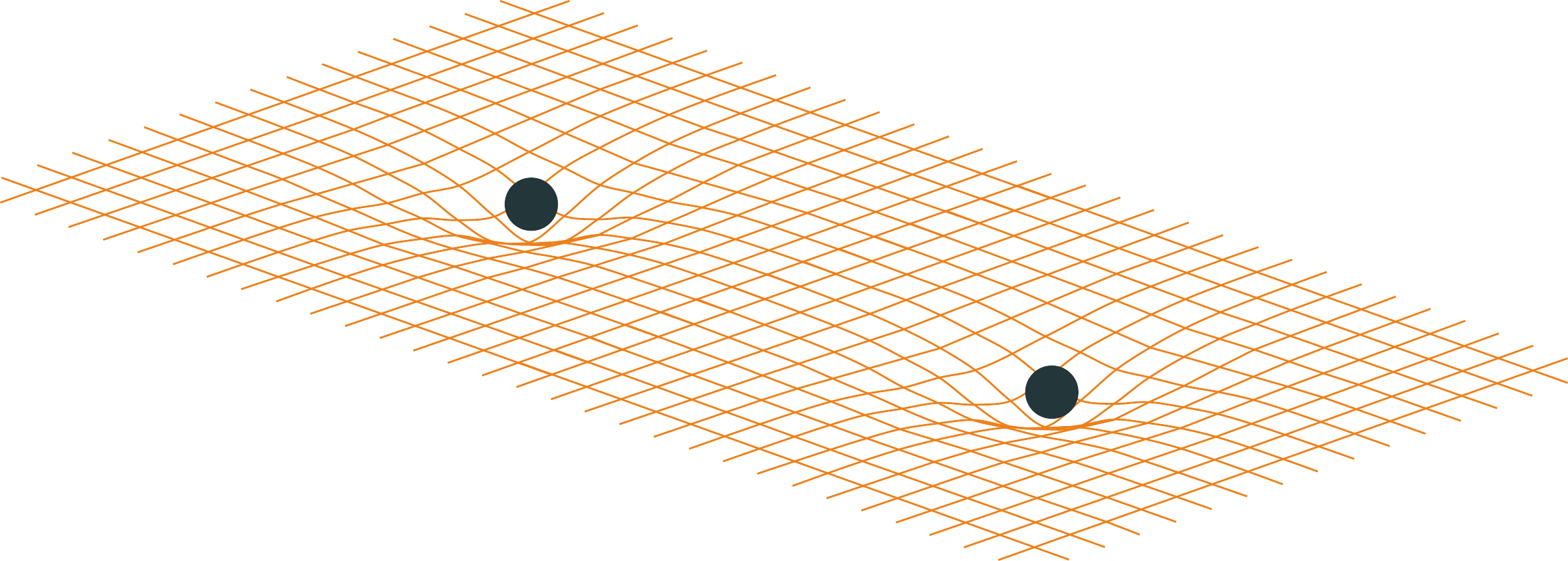}
    \includegraphics[width=0.32\textwidth]{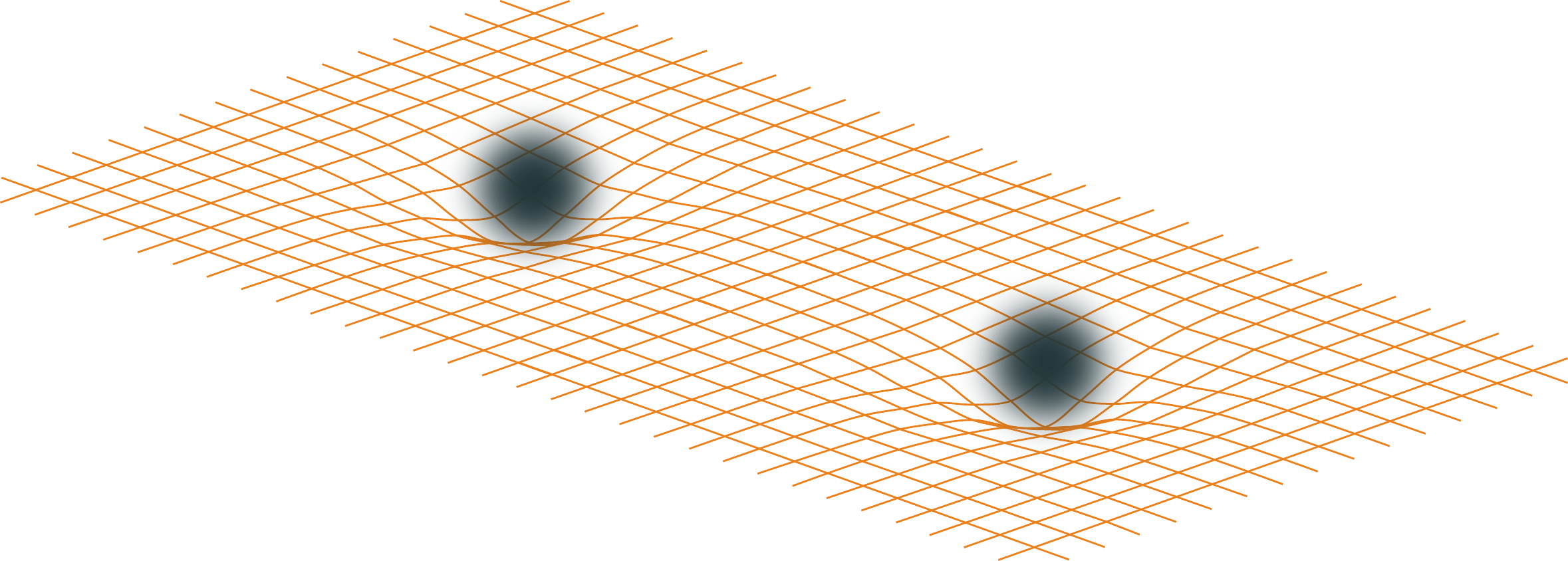}
    \includegraphics[width=0.32\textwidth]{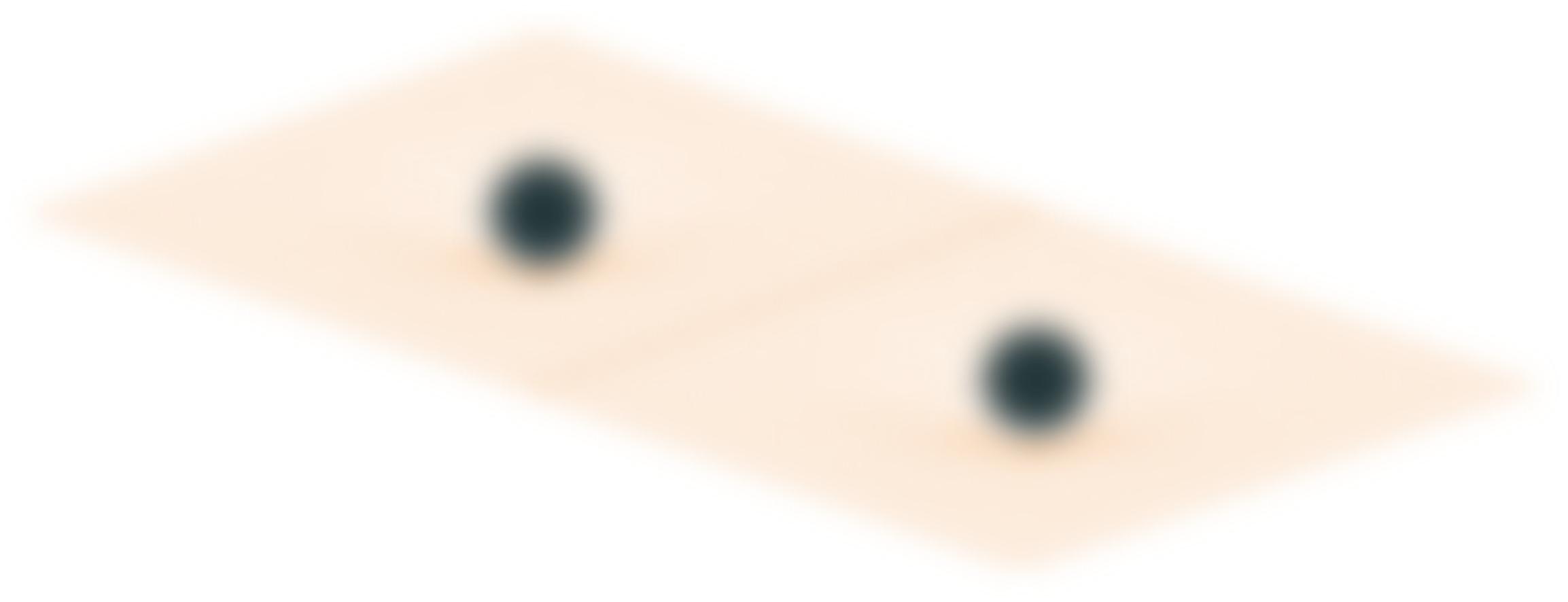}
    \caption{In a classical theory of gravity (left), matter and spacetime are classical. In a semiclassical theory of gravity (center), matter is quantum but there is a single classical spacetime; all particles feel the same. Finally, in a fully quantum theory of gravity (right), matter and spacetime are quantum. While the need for having a quantum description of matter is experimentally demonstrated, it is so far unclear if this is needed for spacetime.}
    \label{fig:3options}
\end{figure}

Semiclassical gravity is sometimes used in a rather restrictive sense, to mean QFTCST alone --that is without any backreaction of matter on spacetime--, to mean the specific M{\o}ller-Rosenfeld backreaction proposal, or finally to designate certain classes of \emph{approximations} to a more fundamental quantum theory of gravity.
Rather, I want to consider a potentially fundamental theory in which quantum matter curves a classical spacetime (in whatever way) and in which matter moves in this spacetime following (at least approximately) QFTCST\footnote{This latter requirement is used to prevent labelling ``semiclassical'' a Bohmian quantum theory of gravity. In such a theory, there would be a well defined trajectory for the metric, but the dynamics for the matter part would not be given by QFTCST with the Bohmian metric.}. Is this possible? Can it be made consistent? If so, is it already falsified or soon falsifiable?

\subsection{In the Newtonian limit}
The difference between a fully quantum theory of gravity and semiclassical gravity is perhaps best illustrated in the Newtonian limit. Consider the Schr\"odinger equation for a matter state $\ket{\psi_t}$:
\begin{equation}
    i \hbar \frac{\upd}{\upd t} \ket{\psi_t} = \left(\hat{H}_0 + \hat{V} \right) \ket{\psi_t}
\end{equation}
where $\hat{H}_0$ is the Hamiltonian in the absence of gravity and $\hat{V}$ is the potential of gravitational origin. If gravity is fully quantum, then $\hat{V}$ is simply a pair potential, in the same way the Coulomb force is obtained in the non-relativistic limit of Quantum Electrodynamics (QED). On the other hand, for a semiclassical theory, all particles see the same classical (here scalar) field $\Phi(\xb)$. Namely:
\begin{align}
   \textsf{Quantum gravity} ~~~ \hat{V} &= -\frac{G}{2}  \int_{\mathds{R}^3}\upd \xb  \int_{\mathds{R}^3}\upd \yb \, \frac{\hat{M}(\xb) \hat{M}(\yb)}{|\xb-\yb|}\label{eq:fullquantum}\\
    \textsf{Semiclassical gravity} ~~~\hat{V} & = \int_{\mathds{R}^3} \upd \xb \; \Phi(\xb) \, \hat{M}(\xb) \label{eq:semiclassical}
\end{align}
where $\hat{M}(\xb)$ is the mass density operator, which reads \eg
\begin{equation}
\hat{M}(\xb) = \sum_{\ell=1}^n m_\ell\, a_\ell^\dagger(\xb) \, a_\ell (\xb),
\end{equation}
for $n$ different species of particles of mass $m_\ell$, with $a_\ell^\dagger(\xb)$ and $a_\ell(\xb)$ the local creation and annihilation operators of particles of type $\ell$. For a semiclassical theory, equation \eqref{eq:semiclassical}, which is uncontroversial, only tells how matter moves once the gravitational field is given. One needs to complement it with a specification for backreaction, which means identifying what object $M(\xb)$ sources the field $\Phi$ in the Poisson equation:
\begin{equation}
    \nabla^2 \Phi(\xb) = 4 \pi G M(\xb).
\end{equation}
In this context, defining a theory of semiclassical gravity requires providing a prescription for $M(\xb)$ and typically its dependence on the quantum state (or on extra variables): this is the non-trivial part. However, whatever choice one makes for the $M(\xb)$ and thus $\Phi$, it is not possible to recover the quantum evolution \eqref{eq:fullquantum} from the semiclassical prescription \eqref{eq:semiclassical}, as it would require an operator valued field $\Phi\rightarrow \hat{\Phi}/{2}$.\footnote{This factor of $1/2$ is often missed in the literature where the semiclassical approach is compared with the purely quantum one. One does not get the correct quantum prescription simply with the naive procedure of putting hats in the Poisson equation and using \eqref{eq:semiclassical}. One needs to take into account the field Hamiltonian, which is not zero in the non-relativistic limit, and which precisely contributes to $- 1/2$ times what the matter-field coupling yields, giving the overall $1/2$ in \eqref{eq:fullquantum}.}

\section{The standard approach and its problems}

We now discuss the standard approach to semiclassical gravity and its legitimate shortcomings. 

\subsection{M{\o}ller-Rosenfeld semiclassical gravity}
To source the gravitational field out of quantum matter, one needs a prescription to extract some classical quantity out of something quantum. A natural option, historically due to M\o ller and Rosenfeld \cite{moller1962,rosenfeld1963}, is to use quantum expectation values to turn operators into scalars and use the result as source in Einstein's equation, \ie to posit that:
\begin{equation}\label{eq:mr}
R_{\mu\nu} -\frac{1}{2} R g_{\mu \nu} =8\pi G \, \langle\psi|\hat{T}_{\mu\nu}|\psi\rangle.
\end{equation}
Let us note already that it is non trivial to make sense of this equation in the general case. One technical difficulty, among others, is that the r.h.s. is generically divergent, even for the vacuum, and needs to be renormalized \cite{wald1978,wald1994}. The technical difficulties of the fully relativistic setting may have overshadowed the conceptual issues that can be seen already in the Newtonian limit. In my opinion, these are the most serious difficulties and the ones that most urgently need to be addressed if one wants to explore the option that gravity is not fully quantum.

\subsection{Newtonian limit}
If \eqref{eq:mr} is taken as fundamental\footnote{The Poisson equation \eqref{eq:poisson} is only a good non-relativistic approximation of \eqref{eq:mr} if the latter is taken to be fundamental \cite{bahrami2014}. Naturally, if \eqref{eq:mr} is aimed to provide a mean-field approximation of a more fundamental theory of gravity, then the Poisson equation \eqref{eq:poisson} for a system of a few particles will be a very poor approximation of the full quantum description \cite{anastopoulos2014}, but this is not the scenario we discuss here.}, then its non-relativistic limit is simply given by the Poisson equation \cite{giulini2012}, where field degrees of freedom are slaved to matter:
\begin{equation}\label{eq:poisson}
\nabla^2 \Phi(\xb,t)=4\pi G\, \bra{\psi_t} \hat{M}(\xb) \ket{\psi_t}.
\end{equation}
Once this classical field is given, which is the non-trivial part, it acts on quantum matter in the standard way, as an external field:
\begin{equation}
\frac{\upd}{\upd t} \ket{\psi} = - \frac{i}{\hbar}\left( H_0 + \int \upd \xb\,\Phi(\xb,t) \hat{M}(\xb) \right)\ket{\psi_t}.
\end{equation}
Inverting the Poisson equation \eqref{eq:poisson}, to replace the field by its matter source, then yields the celebrated Schr\"odinger-Newton equation \cite{diosi1984,bahrami2014}:
\begin{equation}\label{eq:SN}
\frac{\upd}{\upd t} \ket{\psi_t} =-\frac{i}{\hbar}H_0\ket{\psi_t} + i \, \frac{G}{\hbar} \int \upd \xb\,\upd \yb\,\frac{\bra{\psi_t} \hat{M}(\xb)\ket{\psi_t}\, \hat{M}(\yb)}{|\xb-\yb|} \ket{\psi_t}.
\end{equation}
For a single (otherwise free) particle of wave function $\psi$, it takes the perhaps more familiar form:
\begin{equation}
\frac{\upd}{\upd t} \psi(t,\xb) = \frac{i}{2\hbar m}\nabla^2\psi(t,\xb) + i\, \frac{Gm^2}{\hbar} \int \upd \yb\, \frac{|\psi(\yb,t)|^2}{|\xb-\yb|} \,\psi(t,\xb),
\end{equation}
which is cubic in the wavefunction.
With such an evolution, a single particle weakly interacts with itself.

\subsection{The problem of non-linear back-reaction}
The deterministic non-linear character of the Schr\"odinger-Newton brings well known difficulties \cite{gisin1989,gisin1990,polchinski1991,bassi2015} which are hard to bypass.
It is worth discussing these conceptual problems because arguments against semiclassical gravity often leverage them in one way or another.

Discussing the consequences (even only operational) of a non-linear modification of the Schr\"odinger equation is not as easy as one may think. Indeed, non-linearities break the empirical equivalence ``for all practical purposes'' between decoherence and collapse. Consequently different interpretations of quantum theory make different predictions and should be discussed separately.

The first possibility is to consider an interpretation of quantum mechanics in which there is no fundamental collapse process, such as the Many-World interpretation or de Broglie-Bohm (dBB) theory. In such approaches, the problem of the Schr{\"o}dinger-Newton equation is one of \emph{empirical inadequacy}. Indeed, the non-linear gravitational term couples even fully orthogonal branches of the universe wave-function. Decoherence no longer achieves the decoupling role it has in standard linear quantum mechanics: Schr\"odinger's dead and live cat(s), even fully decohered with the environment, attract each other\footnote{The initial hope was that the manifest clash with our experience this implies would be avoided because the non-linear addition would also dynamically collapse macroscopic superpositions precisely in the way one would want, according to the Born rule. Unfortunately, this is not the case \cite{bahrami2014}.}.  Hence for interpretations of quantum mechanics without fundamental collapse, the theory is well defined but makes predictions that seem completely at odds with our experience.

\begin{figure}
    \centering
    \includegraphics[width=0.34\textwidth]{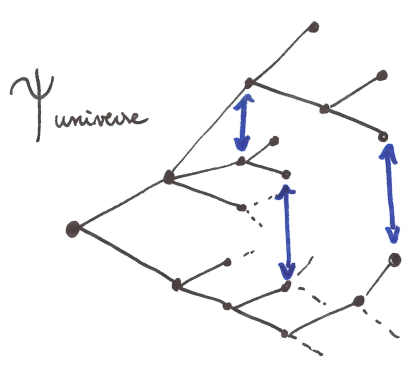}~~~~~~~~~~~~~~~
    \includegraphics[width=0.45\textwidth]{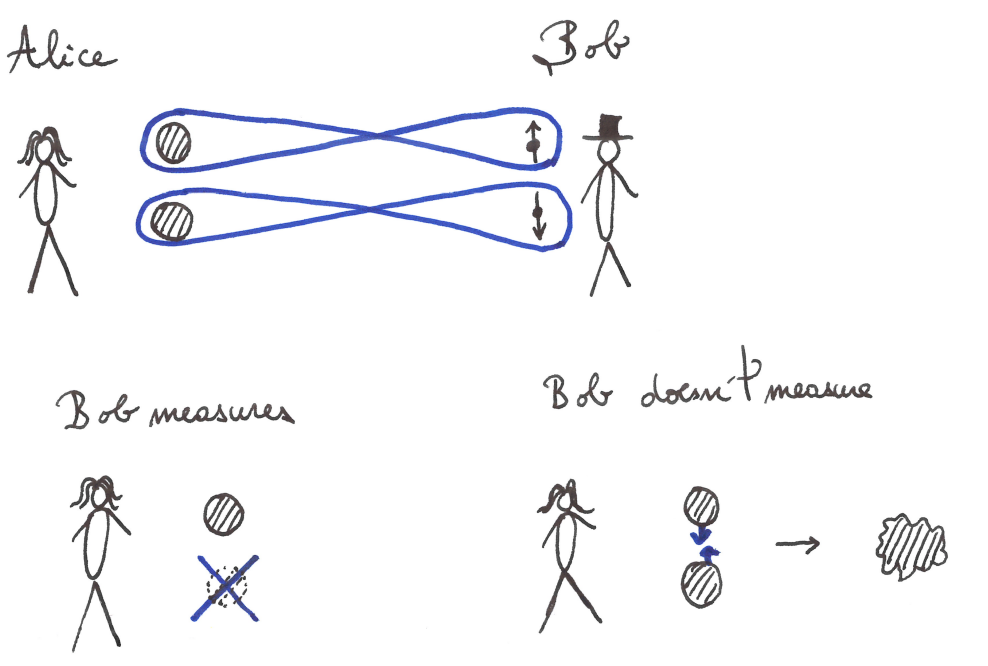}
    \caption{\textit{Left}: without collapse, orthogonal branches of the universe wavefunction attract each other via the non-linear gravitational interaction. \textit{Right}: Alice and Bob share an entangled pair with a massive superposition on Alice's side. In linear quantum mechanics, the statistics on Alice's side do not depend on what Bob does on his side. In a non-linear theory, Alice can know, via measurements on \emph{her} side, if Bob carried a measurement on his side. Note that this information is not transmitted by gravity from Bob to Alice. Rather, it is the local non-linearity of gravity that allows to witness the non-local collapse of the wave function.}
    \label{fig:alicebob}
\end{figure}

The second option is to have an interpretation in which collapse is fundamental and happens (not necessarily exclusively) in measurement situations. In that case, the difficulty is a bit more subtle and lies in the breakdown of the statistical interpretation of the quantum state. Consider a situation in which Alice and Bob share an Einstein-Podolski-Rosen (EPR) pair, with a massive superposition on Alice's side (see Fig. \ref{fig:alicebob}). If the wave-function collapses upon measurement on Bob's side, then Alice's mass is localized in a given position and nothing happens to it. If Bob carries no measurement and the wave-function does not collapse, Alice's mass stays superposed and evolves with the non-linear self gravitation term. Alice can easily distinguish the two situations, e.g. by making a measurement in the region where most the mass would lie in the second situation. Not only would this generically allow faster than light signalling (at least in one frame), the Born rule would break down when used even on well isolated subsystems. Alice's density matrix is insufficient to determine statistics on her side. 

In practice, the kind of massive superpositions required to make this example clean may very well be suppressed by an intrinsic collapse process, but the conclusion remains that it would in general be impossible to use the statistical interpretation of the state vector and that there would  be residual faster than light signalling effects.

\subsection{Conclusion on standard semiclassical gravity}
The standard (M{\o}ller-Rosenfeld) approach to semiclassical gravity is not completely killed by the previous arguments. Especially in the second case, for interpretations of quantum mechanics with an objective collapse process, one may hope that violations of the statistical interpretation of the quantum state do not amplify from micro to macro, remaining generically small and under control. Further, faster-than-light effects could be so small that no practical consequences would ensue, and restricted to a single frame such that no theoretical inconsistency could be derived. Hence, it is technically possible to still defend this approach provided one is ready to pay the price of an auxiliary collapse process and of serious practical complications (the Born rule becoming only approximately valid, without any good control on the approximation).

Nonetheless, I think it is fair to say that if this were the only representative of semiclassical theories, the case for fundamental semiclassical gravity would be rather slim. However, one should remember that, while simple, the \emph{choice} of making expectation values the classical stuff that sources the gravitational field is not unique nor even natural. At a heuristic level, the M{\o}ller-Rosenfeld prescription assumes that $|\psi|^2$ is what is ``real'', that it is the tangible thing that bends spacetime. But we are free to explore other options.

\section{Discrete models restoring linearity}

\subsection{Basic idea}

A natural thing to do if one wants to get something classical from quantum matter is to go to an interpretation of quantum mechanics where there is such a thing as classical stuff by construction. For example, one could use Bohmian particles as classical sources of the gravitational field. In fact, this has been tried by Struyve \cite{struyve2015,struyve2017}. However, while such a theory could behave in a better way than the previous approach, and importantly not require an additional collapse process to make sense, it still suffers from the breakdown of the statistical interpretation and the theoretical possibility to send information faster-than-light. In that case, these unappealing features emerge from the breakdown of equivariance which usually insures that the distribution of Bohmian particles is provided by the Born rule.

Hence it seems that rather generically, when one extracts a classical quantity from quantum matter, through the wave-function directly (with quantum expectation values) or through additional variables (Bohmian positions), one breaks the simplicity of the formalism, i.e. the ability to extract predictions from a simple statistical prescription.

However, this difficulty can be avoided and it seems at present that there is essentially only one simple way to do it. The intuition comes from (orthodox) quantum measurement and feedback. In a lab, an experimentalist measures a quantum systems, obtains a certain result, processes it, and applies a control pulse depending on this result. This complete procedure of measurement and feedback does not break the formalism as the latter was precisely constructed to describe such situations. Further, the measurement result is a tangible thing, which after some (classical) processing is used to act on the quantum state. Measurement and feedback thus allow the consistent coupling of classical and quantum systems. This intuition can be used to construct (potentially) fundamental semiclassical theories that are without conceptual problems \cite{diosi1998,halliwell1999,diosi2017}. In the context of gravity, this intuition was first explored in a toy model by Kafri \textit{et al.} \cite{kafri2014} where it was called ``classical channel gravity''.  

Of course, this measurement and feedback intuition is only used to construct a model with good properties. It should not be understood ``ontologically'' as meaning there are actual detectors carrying the measurement task. 

\subsection{Intuition in the discrete}

Let us consider a rather general weak measurement with $n$ outcomes, parameterized by $n$ matrices $\hat{N}_k$ which verify as only constraint $\sum_k \hat{N}_k^\dagger \hat{N}_k = \mathds{1}$. Such a measurement acts on a quantum state in the following way:
\begin{equation}\label{eq:povm}
    \ket{\psi} \xrightarrow{\text{measurement}} \ket{\psi_{k}}=\frac{\hat{N}_k \ket{\psi}}{\sqrt{\bra{\psi} \hat{N}_k^\dagger \hat{N}_k \ket{\psi}}},
\end{equation}
with probability $p_k=\bra{\psi} \hat{N}_k^\dagger N_k \ket{\psi}$ where $k$ labels the measurement outcome. This is the first step of any measurement + feedback scheme.

Now, ``feeding back'' the result $k$ just amounts to apply a $k$-dependent unitary operator $\hat{U}_k$ on the state. That is, for a measurement immediately followed by a feedback (or neglecting any intrinsic system dynamics happening in the meantime), one has:
\begin{equation}\label{eq:discretemeasurement}
     \ket{\psi} \xrightarrow[+\; \text{feedback}]{\text{measurement}} \ket{\psi_k^\text{f.b.}}= \hat{U}_k \ket{\psi_k}=\hat{U}_k\frac{\hat{N}_k \ket{\psi}}{\sqrt{\bra{\psi} \hat{N}_k^\dagger \hat{N}_k \ket{\psi}}},
\end{equation}
with the same probability $p_k$. Notice that the full evolution is still a legitimate generalized measurement with the substitution $\hat{N}_k\rightarrow \hat{B}_k= \hat{U}_k \hat{N}_k$. 

By construction, this measurement and feedback scheme does not break the standard formalism despite the fact that the measurement step is non-linear. Indeed this non-linearity vanishes upon averaging over measurement results:
\begin{equation}\label{eq:discretefeedback}
   \rho^\text{f.b.} \equiv \mathds{E}\left[\ket{\psi^\text{f.b.}}\bra{\psi^\text{f.b.}}\right] = \sum_{k=1}^n p_k  \frac{\hat{B}_k \ket{\psi}\bra{\psi} \hat{B}_k^\dagger}{\bra{\psi} \hat{N}_k^\dagger \hat{N}_k \ket{\psi}} \equiv \sum_{k=1}^n \hat{B}_k\underset{\rho}{\underbrace{\ket{\psi}\bra{\psi}}}\hat{B}_k^\dagger,
\end{equation}
where $\mathds{E}[\,\cdot\,]$ denotes the statistical average. Any modification of the measurement rule \eqref{eq:discretemeasurement}, which specifies how the classical variable $k$ is extracted from the quantum state, would generically preclude this cancellation of non-linearities. Unless there is a particular reason why linearity should hold, it typically does not.

\subsection{Simplest discrete model for gravity}
With this simple understanding of feedback, one can already construct a toy model of semiclassical gravity with discrete Poisson distributed measurement and feedback events \cite{tilloy2017grw}.

Let us consider for convenience $N$ distinguishable particles with coordinates $\{\xb_1,\cdots,\xb_N\}$, masses $\{m_1,\cdots,m_N\}$, and without spin. The $N$-particle system is described by the wave-function $\psi_t(\xb_1,\cdots,\xb_N)$ with $\xb_\ell\in \mathds{R}^3$. Our objective is to implement a consistent semiclassical gravitational force between them. We consider the following weak position measurement of particle $\ell$:
\begin{equation}\label{eq:jump}
    \psi_t \longrightarrow \frac{\hat{L}_\ell(\xb_f) \psi_t}{\|\hat{L}_\ell(\xb_f) \psi_t\|},
\end{equation}
with the operator
\begin{equation}
\hat{L}_\ell(\xb_f)=\frac{1}{(\pi r^2_c)^{3/4}} \mathrm{e}^{-(\hat{\xb}_\ell-\xb_f)^2/(2 r^2_c)},
\end{equation}
where $\hat{\xb}_\ell$ is the position operator associated to particle $\ell$. The probability for the outcome $\xb_f$ is $P(\xb_f)=\|\hat{L}_k(\xb_f) \psi_t\|^2$. This is as just before in eq. \eqref{eq:discretemeasurement} with discrete indices labeling results promoted to continuous position variables $k\rightarrow \xb_f$ and $\hat{N}_k \rightarrow \hat{L}_\ell(\xb_f)$. The parameter $r_c$ encodes the sharpness of the position measurement.

We now assume that each particle can be hit by this spontaneous weak measurement process independently of the others and with a distribution uniform in time with intensity $\lambda$. The astute reader will have noticed that this random dynamics for the quantum state is nothing but the Ghirardi-Rimini-Weber (GRW) model \cite{ghirardi1986,bassi2003,bassi2013}, the simplest dynamical reduction model. However, its measurement interpretation is fundamental in that it guides us to the proper choice of variables to use as sources for the gravitational field.

Now assume that a collapse (or weak measurement) with center (or result) $\xb_f$ has hit particle $\ell$ at time $t_f$. We now take this space-time event (or ``flash'') $(t_f,\xb_f)$ as source of the gravitational field\footnote{This possibility was mentioned, although not explored further, by Derakhshani \cite{derakhshani2014}.}:
\begin{equation}\label{eq:flashingsource}
\nabla^2 \Phi(\xb,t) = 4\pi G m_{\ell} \lambda^{-1} \delta(t-t_f) \delta(\xb-\xb_f).
\end{equation}
More generally, one could replace these Dirac distributions by a generic causal spacetime smearing function $f$ such that:
\begin{equation}
    \int_{\mathds{R}^4} \upd t \,\upd \xb  \, f(t,\xb) = 1~~~~~~~ f\geq 0 ~~~~~~~ \forall t<0, ~~f(t,\xb) = 0.
\end{equation}
Since the measurement step has resolution $r_c$, it would \eg also make sense to have $f$ spatially smear the point source with a Gaussian of width $r_c$. The objective here is just to provide the simplest model and thus we will just consider punctual sources. The gravitational potential created by this event is:
\begin{equation}
    \hat{V} = \int \upd \xb \Phi(\xb) \hat{M}(\xb),
\end{equation}
where the mass density operator is simply  $\hat{M}(\xb) = \sum_{\ell=1}^N m_\ell \, \delta(\xb-\hat{\xb}_\ell)$ in this first quantized representation. Inverting the Poisson equation \eqref{eq:flashingsource} yields:
\begin{equation}
    \hat{V}(t) = -\delta(t-t_f)\, G \lambda^{-1} \sum_{\ell'=1}^N   \frac{m_\ell m_{\ell'}}{|\xb_f-\hat{\xb}_{\ell'}|}.
\end{equation}
This time-singular potential may be worrying at first sight, but once put in the Schr\"odinger equation it simply implements a unitary transformation:
\begin{equation}
\hat{U}_\ell(\xb_f)=\exp\left(i \,\frac{G}{\lambda \hbar}\sum_{\ell'=1}^N \frac{m_\ell m_{\ell'}}{|\xb_f-\hat{\xb}_{\ell'}|}\right),
\end{equation}
which happens immediately \emph{after} the collapse. As before, it just changes the collapse operators $\hat{L}_\ell(\xb_f) \rightarrow \hat{C}_\ell(\xb_f) = \hat{U}_\ell(\xb_f) \hat{L}_\ell(\xb_f)$. This fully specifies the model. In a nutshell, the position of each particle is weakly measured and the measurement result is used to create a unitary gravitational kick. In the language of the GRW model, we just made one possible choice of local beables  \cite{bell1976} (or primitive ontology \cite{allori2014,allori2015,tumulka2011}), the \emph{flashes}, the source of the gravitational field \cite{tilloy2017grw}.

The master equation for this model is very simple and reads:
\begin{equation}\label{eq:master2}
\frac{\upd}{\upd t} \rho_t = -\frac{i}{\hbar}[\hat{H}_0,\rho_t] + \lambda \sum_{\ell=1}^N\left\{ \int_{\mathds{R}^3}   \hat{C}_\ell(\xb_f)\rho_t \hat{C}^\dagger_\ell(\xb_f)  \, \upd \xb_f -  \rho_t \right\},
\end{equation}
where $\hat{H}_0$ is just the standard Schr\"odinger Hamiltonian containing all interactions but gravity. Importantly, it is linear, and thus the statistical intepretation of the density matrix survives. This model has a slightly more complicated phenomenology than continuous alternatives, and the fact that gravity acts through punctual kicks creates a new length scale. Indeed, assuming all the masses are the same $m_k=m$:
\begin{equation}
    \hat{U}_\ell(\xb_f) = \exp\left( i \sum_{\ell'=1}^N \frac{r_G}{|\xb_f-\hat{\xb}_{\ell'}|}\right),
\end{equation}
with $r_G = G m^2/(\hbar \lambda)$. For $\lambda=10^{-16}\mathrm{s}^{-1}$ (historical value of $\lambda$ for the GRW model that is not yet experimentally falsified) we get $r_G\simeq1.8\times 10^{-14}\mathrm{m}$ for protons and $r_G \simeq 5.3\times 10^{-21} \mathrm{m}$ for electrons. In both cases, $r_G\ll r_c$, the smearing length of the measurement. If we consider interparticle distances $r$ much larger than $r_G$ we can expand the unitary operator to first order in $r_G/r$ and we obtain:
\begin{equation}\label{eq:master3}
    \frac{\upd}{\upd t} \rho_t\simeq -\frac{i}{\hbar}[\hat{H}_0 + ,\rho_t] -\frac{i}{\hbar}\hat{V}_\text{eff}\cdot \rho_t+ \lambda \sum_{\ell=1}^N\left\{ \int_{\mathds{R}^3}   \hat{L}_\ell(\xb_f)\rho_t \hat{L}^\dagger_\ell(\xb_f)  \, \upd \xb_f -  \rho_t \right\},
\end{equation}
with 
\begin{equation}
    \hat{V}_\text{eff} \cdot \rho= \sum_{\ell=1}^N\sum_{\ell'=1}^N\int_{\mathds{R}^3} \upd \xb_f \left[-\frac{Gm_\ell m_{\ell'}}{|\xb_f-\hat{\xb}_{\ell'}|}, \hat{L}_\ell(\xb_f) \rho\hat{L}^\dagger_\ell(\xb_f)\right],
\end{equation}
which looks a lot like the purely quantum pair potential up to some smearing around $\rho$. The way the expected pair potential is approximately recovered will be clearer for continuous models, but it can already be seen here for a simple situation. Consider $N$ particles separated from each other by an average distance $r\gg r_c$. Writing $\rho^{(1)} =\tr_{N-1}\left[\rho\right]$ the reduced density matrix of the first particle, one has \cite{tilloy2017grw}:
\begin{equation}
\tr_{N-1}\left[\hat{V}_\text{eff} \cdot \rho\right] \simeq \tr_{N-1}\left( \left[-\sum_{\ell=2}^N \frac{Gm_1 m_\ell}{|\hat{\xb}_1-\hat{\xb}_\ell|},\rho \right]\right).
\end{equation}
Further assuming that the $N-1$ particles traced over are localized over a distance much smaller than their distance to particle $1$ yields:
\begin{equation}
\tr_{N-1}\left[\hat{V}_\text{eff} \cdot \rho\right] \simeq \left[-\sum_{\ell=2}^N \frac{Gm_1 m_\ell}{|\hat{\xb}_1-\xb_\ell|},\rho^{(1)} \right],
\end{equation}
which is precisely the potential one would have put by hand in a scenario where a quantum particle interacts with $N-1$ fixed classical particles.

In this model, gravity behaves in a rather peculiar way, through discrete kicks. Yet, it is not obviously falsified by experiments probing gravity in the Newtonian limit \cite{tilloy2017grw}.

\section{Continuous models restoring linearity}

The previous model is easy to define, and the reason why it preserves linearity at the master equation level is particularly understandable and straightforward to verify. However, its phenomenology is not completely trivial, and the fact that gravity acts through punctual kicks makes everything more complicted. This motivates the discussion of continuous models, which were actually introduced first.

\subsection{Continuous measurement and feedback}
When infinitely weak measurements are repeated infinitely frequently, one can obtain a continuous measurement described by a diffusive stochastic differential equation \cite{jacobs2006, wiseman2009}. The real-time result (or signal) from the continuous measurement of a self-adjoint operator $\mathcal{O}$ reads:
\begin{equation}\label{eq:signal}
    \mathscr{S}_\mathcal{O}(t) = \langle \mathcal{O}\rangle_t + \eta_t,
\end{equation}
where $\langle \mathcal{O}\rangle_t = \bra{\psi_t}\mathcal{O}\ket{\psi_t}$ and $\eta$ is white noise:
\begin{equation}
    \mathds{E}\left[\eta_t\eta_s\right] = \frac{1}{\gamma} \delta(t-s) ~~~~~~~~\mathds{E}[\eta_t] = 0,
\end{equation}
where $\gamma$ encodes the measurement strength.
The signal given by equation \eqref{eq:signal} is the equivalent of the result $k$ and its probability distribution $P(k)=\bra{\psi}N^\dagger_k N_k\ket{\psi}$ in the discrete.

The corresponding non-linear backaction on the quantum state reads:
\begin{equation}\label{eq:nonitoback}
    \frac{\upd}{\upd t} \ket{\psi}_t= \left[ \frac{\gamma}{2} \left(\mathcal{O} - \langle \mathcal{O}\rangle_t\right)\eta_t -\frac{\gamma}{4}\left(\mathcal{O} - \langle \mathcal{O}\rangle_t\right)^2\right] \ket{\psi_t}.
\end{equation}
This is a stochastic differential equation with multiplicative white noise, and it should be understood in the It\^o convention. In this context, one needs to be particularly careful with stochastic calculus conventions as corrections to the standard differentiation rules coming from It\^o's formula are necessary to get the physically correct results. This has sometimes led to confusions in the literature. Hence, to mark the It\^o convention, and at the risk of seeming overly pedantic, I now switch to the It\^o notation and rewrite equations \eqref{eq:signal} and \eqref{eq:nonitoback}:
\begin{align}
    \upd R_t &= \langle \mathcal{O}\rangle_t\,\upd t + \upd W_t~~~ ``=" ~~\mathscr{S}_\mathcal{O}(t) \, \upd t \label{eq:itonotationsig}\\
   \upd\ket{\psi}_t&= \left[ \frac{\gamma}{2} \left(\mathcal{O} - \langle \mathcal{O}\rangle_t\right) \, \upd W_t -\frac{\gamma}{4}\left(\mathcal{O} - \langle \mathcal{O}\rangle_t\right)^2 \upd t \right] \ket{\psi_t},\label{eq:itonotationback}
\end{align}
with the heuristic It\^o rule $\upd W_t \upd W_t =\gamma^{-1} \,\upd t$. For all practical purposes, one obtains the It\^o formula for differentiation simply by expanding all differentials to second order in $\upd W$ and using this heuristic rule. For example, one can derive the stochastic master equation for the (pure) density matrix $\rho_t = \ket{\psi_t}\bra{\psi_t}$ writing:
\begin{align}
    \upd \rho_t &= \upd \ket{\psi_t} \; \bra{\psi_t} + \ket{\psi_t} \,\upd \bra{\psi_t} + \underset{\text{It\^o correction}}{\underbrace{\upd \ket{\psi_t} \upd \bra{\psi_t}}}\\ &=-\frac{\gamma}{8}\left[\mathcal{O},\left[\mathcal{O},\rho_t\right]\right] \, \upd t + \frac{\gamma}{2} (\mathcal{O}\rho_t + \rho_t \mathcal{O} - 2 \langle\mathcal{O}\rangle_t \rho_t) \,  \upd W_t,
\end{align}
the latter containing exactly the same information as \eqref{eq:itonotationback}.
Because $\rho_t$ is what shows up when computing expectation values, one often works with it directly instead of the state $\ket{\psi_t}$ even if $\rho_t$ is of rank 1. An important property of stochastic differential equations in the It\^o convention is that in integrals of the form $\int f_t \upd W_t$, $f_t$ and $\upd W_t$ are uncorrelated. As a result, taking the average of a stochastic differential equation simply removes the noise term. Writing $\bar{\rho}=\mathds{E}\left[\rho\right]$ we thus get the master equation
\begin{equation}\label{eq:mastermeas}
    \frac{\upd }{\upd t}\rhoav= -\frac{\gamma}{8}\left[\mathcal{O},\left[\mathcal{O},\rhoav\right]\right]. 
\end{equation}
As expected, this master equation is linear: continuous measurements, just like discrete ones, bring no non-linearities at the master equation level.

Let us now go back to the stochastic master equation without averaging and introduce a feedback term depending on the real-time signal. Namely, let us heuristically consider the addition of a potential proportional to the signal
\begin{equation}\label{eq:feedbackinformal}
    \hat{V}^\text{f.b.}(t) ~~``=" ~~ \mathscr{S}_\mathcal{O}(t)\,\hat{K} 
\end{equation}
in the Schr\"odinger equation, where $\hat{K}$ is self-adjoint. Because this potential contains white noise, one needs to be careful to implement it. Importantly, this potential does not act at the exact same instant as the measurement from which the result is extracted, but infinitesimally \emph{after}. This is analogous to the discrete situation where the feedback unitary $\hat{U}_k$ is applied just after the measurement operator $\hat{N}_k$. Hence we have to translate the informal procedure \eqref{eq:feedbackinformal} into the rigorous prescription
\begin{equation}\label{eq:feedbackformal}
    \rho_t + \upd \rho_t^\text{f.b.} =\exp(-i \hat{K}\upd R_t/\hbar) (\rho_t + \upd \rho_t) \exp(i \hat{K}\,\upd R_t/\hbar),
\end{equation}
where $\upd \rho_t$ is the infinitesimal evolution of $\rho_t$ given by pure continuous measurement \eqref{eq:itonotationback}. Expanding all terms up to second order in $\upd W$ and using the heuristic It\^o rule yields
\begin{equation}\label{eq:feedbacksme}
    \upd \rho_t^\text{f.b.} = \upd \rho_t - \frac{i}{\hbar}[\hat{K},\rho_t]\, \upd W_t - \frac{i}{2} \left[\hat{K},\left\{\mathcal{O}, \rho_t\right\}\right]\, \upd t - \frac{1}{2\,\gamma\hbar^2}\left[\hat{K}\left[\hat{K},\rho_t\right]\right] \,\upd t.
\end{equation}
Note that one gets the same stochastic master equation with the more physical procedure of regulating the signal with a low pass filter first, and subsequently sending the frequency cutoff of the filter to infinity (or equivalently the associated smearing timescale to zero) \cite{wiseman2009}. One may again average over the measurement randomness to get the linear master equation:
\begin{equation}\label{eq:singleoperatorfeedback}
    \frac{\upd }{\upd t}\rhoav= - \frac{i}{2\hbar} \left[\hat{K},\left\{\mathcal{O}, \rhoav\right\}\right] -\frac{\gamma}{8}\left[\mathcal{O},\left[\mathcal{O},\rhoav\right]\right] - \frac{1}{2\gamma \hbar}\left[\hat{K}\left[\hat{K},\rhoav\right]\right].
\end{equation}
At the master equation level, this feedback has thus added a new decoherence term $\propto [\hat{K},[\hat{K},\rho]]$ and a so far hard to interpret term $\propto -i\left[\hat{K},\left\{\mathcal{O}, \rhoav\right\}\right]$ which effectively acts like a potential up to an additional non-trivial dissipation.
We will need a generalization of this procedure to the continuous simultaneous measurement of $n$ self-adjoint operators $\mathcal{O}^j$. In that case, a fairly general consistent continuous measurement model can be written:
\begin{align}
    \upd R_t^j &= \langle \mathcal{O}^j\rangle_t + \upd W_t^j\\
    \upd \rho_t&=-\frac{1}{8}\Gamma_{jk}\left[\mathcal{O}^j,\left[\mathcal{O}^k,\rho_t\right]\right]\,\upd t  + \frac{1}{2}\Gamma_{jk} \left(\mathcal{O}^j\rho_t + \rho_t \mathcal{O}^j - 2 \langle\mathcal{O}^j\rangle_t \rho_t \right) \,\upd W_t^k,
\end{align}
with the It\^o rule $\upd W_t^j \upd W_t^k=(\Gamma^{-1})^{kj}\, \upd t$ where $\Gamma$ is a \emph{real} symmetric positive definite matrix and we have used implied summation on repeated indices. Intuitively, in a measurement situation, this $\Gamma$ matrix encodes the correlation of the detectors and its eigenvalues encode the measurement strength. In this more general setup, a general real time feedback prescription reads:
\begin{equation}
    \upd\hat{V}^\text{f.b.}= \hat{K}_j \,\upd R^j_t,
\end{equation}
which yields a stochastic master equation for $\rho_t$ that can be derived in the same way as in equations \eqref{eq:feedbackformal} and \eqref{eq:feedbacksme}. Upon averaging over the measurement randomness, one finally gets a master equation generalizing \eqref{eq:singleoperatorfeedback}:
\begin{equation}\label{eq:mastern}
    \frac{\upd }{\upd t}\rhoav= - \frac{i}{2 \hbar} \left[\hat{K}_j,\left\{\mathcal{O}^j, \rhoav\right\}\right] -\frac{\Gamma_{jk}}{8}\left[\mathcal{O}^j,\left[\mathcal{O}^k,\rhoav\right]\right] - \frac{(\Gamma^{-1})^{jk}}{2\hbar^2}\left[\hat{K}_j\left[\hat{K}_k,\rhoav\right]\right].
\end{equation}
The first term in this master equation, which contains a mix of unitary and dissipative effects, may again seem mysterious. It is simpler to interpret when $\hat{K}_j\otimes\mathcal{O}^j= \mathcal{O}^j\otimes\hat{K}_j$, in which case there is no dissipation and we are left with a pure potential term:
\begin{equation}\label{eq:qpotential}
    - \frac{i}{2\hbar} \left[\hat{K}_j,\left\{\mathcal{O}^j, \rhoav\right\}\right] = -\frac{i}{4\hbar}\left[\left\{\mathcal{O}^j,\hat{K}_j\right\}, \rhoav\right].
\end{equation}
In that case, which is the one we will be considering for gravity, measurement + feedback allows to implement a purely quantum potential at the price of additional decoherence coming from the measurement itself and from the feedback noise. These two sources of decoherence, by design, make sure that the interaction is in fact purely classical despite what the presence of the quantum potential \eqref{eq:qpotential} in the master equation \eqref{eq:mastern} may suggest.

\subsection{General continuous mass density measurement and feedback}
We are now equipped to implement gravity by a continuous measurement of the mass density and feedback of the corresponding gravitational potential. To this end, we start from a continuous measurement model of (regularized) local mass density in every point of space. The corresponding signal reads:
\begin{align}
     &\upd R(\xb,t) = \langle \hat{M}_{r_c}(\xb)\rangle_t + \upd W_t(\xb) ~~``=" ~~    \mathscr{S}_{M_{r_c}(\xb)}(t) \,\upd t \label{eq:masssignal}\\ 
  & \text{with}~~~ \upd W_t(\xb)\upd W_t(\yb) = \Gamma^{-1}(\xb,\yb) \,\upd t.
\end{align}
The corresponding backaction on the state is
\begin{equation}
\begin{split}
    \upd \rho_t =& -\frac{1}{8} \int \upd \xb \upd \yb\; \Gamma(\xb,\yb)  \left[\hat{M}_{r_c}(\xb)\left[\hat{M}_{r_c}(\yb),\rho_t\right]\right]\upd t \\
    &+\frac{1}{2}  \int \upd \xb \upd \yb\; \Gamma(\xb,\yb) \left(\hat{M}_{r_c}(\xb) \rho_t + \rho_t\hat{M}_{r_c}(\xb)- 2\, \langle \hat{M}_{r_c}(\xb)\rangle_t \rho_t\right)\upd W_t(\yb),
\end{split}
\end{equation}
with $\hat{M}_{r_c}(\xb)$ a regularized mass density operator:
\begin{equation}
\hat{M}_{r_c}(\xb) = \left[g_{r_c} * \hat{M}\right](\xb) = g_{r_c}*\sum_{\ell=1}^n m_\ell\, a_\ell^\dagger(\xb) \, a_\ell (\xb),
\end{equation}
where ``$*$'' denotes the convolution product and $g_{r_c}$ is a Gaussian smearing of width $r_c$
\begin{equation}
    g_{r_c}(\xb) = \ e^{-\xb^2/(2r_c^2)}/(\sqrt{2\pi r_c^2})^3.
\end{equation}
This continuous measurement model of the regularized mass density everywhere in space, if taken as fundamental, is simply a continuous collapse model. More precisely, one obtains:
\begin{enumerate}
    \item The Continuous Spontaneous Localization (CSL) model \cite{pearle1989,ghirardi1990} for:
    \begin{equation}
    \Gamma^\text{CSL}(\xb,\yb) = \frac{4 \gamma}{ m_N^2} \times \delta(\xb-\yb),
\end{equation}
where $m_N$ is the mass of a nucleon and $\gamma$ is the collapse ``strength''. It is a rate $\times$ distance$^3$, the corresponding rate is $\lambda_{\text{CSL}}\equiv \gamma/(4\pi r_c^2)^{3/2} $ historically fixed at $\lambda_{\text{CSL}}\simeq 10^{-16}$s$^{-1}$ (the so called ``GRW'' value).
 \item The Di\'osi-Penrose (DP) model \cite{diosi1987,penrose1996} for:
 \begin{equation}\label{eq:DPchoice}
     \Gamma^\text{DP}(\xb-\yb)=\frac{2 G}{\hbar} \times \frac{1}{|\xb-\yb|}.
 \end{equation}
\end{enumerate}
The interest of knowing that, at a mathematical level, this model has a continuous measurement interpretation, is that it allows us to introduce the ``signal'' $\mathscr{S}_{\hat{M}_{r_c}(\xb)}$ \eqref{eq:masssignal}. This is not a natural object to consider in the collapse model context, but it is fundamental to implement a consistent semiclassical interaction.

As in the discrete, we can implement gravity by sourcing the gravitational field from the mass density signal:
\begin{equation}
    \nabla^2 \Phi_t(\xb) = 4 \pi G \mathscr{S}_{M_{r_c}(\xb)}(t).
\end{equation}
As before, this expression is a bit formal because the signal has white noise fluctuations. We nonetheless keep on formally, and invert the Poisson equation to get:
\begin{equation}
    \Phi_t(\xb) = - G \int \upd \yb \; \frac{\mathscr{S}_{M_{r_c}(\xb))}(t)}{|\xb-\yb|}.
    \end{equation}
The corresponding signal dependent feedback potential to introduce in the Schr\"odinger equation thus reads (still formally):
\begin{equation}
    \hat{V}^\text{f.b.} = \int \upd\xb \,  \Phi_t(\xb) \hat{M}_{r_c}(\xb).
\end{equation}
This is equivalent with feeding back  directly the operator $\hat{\Phi}_{r_c}(\xb)$ multiplied by the mass density signal with
\begin{equation}
    \hat{\Phi}_{r_c}(\xb) := - G \int \upd \yb\,  \frac{\hat{M}_{r_c}(\yb)}{|\xb - \yb|}.
\end{equation}
Now more rigorously, we can write the elementary feedback step as in \eqref{eq:feedbackformal}:
\begin{equation}
        \rho_t + \upd \rho_t^\text{f.b.} =\exp(-i \upd \hat{V}^\text{f.b.}/\hbar) (\rho_t + \upd \rho_t) \exp(i \upd \hat{V}^\text{f.b.}/\hbar)
\end{equation}
with
\begin{equation}
     \upd \hat{V}^\text{f.b.}=- G \int \upd \xb \upd \yb\; \frac{\hat{M}_{r_c}(\xb)\,\upd R_t(\yb)}{|\xb-\yb|} = \int \upd \xb  \, \hat{\Phi}_{r_c}(\xb) \, \upd R_t(\xb).
\end{equation}
Following the same steps as before, it yields the master equation for matter
\begin{align}\label{eq:master4}
    \frac{\upd}{\upd t} \rhoav &= -\frac{i}{\hbar}\left[ \hat{V} ,\rhoav \right] -\underset{\text{measurement decoherence}}{\underbrace{\int \!\! \upd \xb \upd \yb \frac{\Gamma(\xb,\yb) }{8} \left[\hat{M}_{r_c}(\xb)\left[\hat{M}_{r_c}(\yb),\rhoav\right]\right]}} -\underset{\text{feedback induced decoherence}}{\underbrace{\int \!\! \upd \xb \upd \yb \frac{\Gamma^{-1}(\xb,\yb) }{2\hbar^2} \left[\hat{\Phi}_{r_c}(\xb)\left[\hat{\Phi}_{r_c}(\yb),\rhoav\right]\right] }}
\end{align}
with the pair potential
\begin{equation}
    \hat{V}=-\frac{G}{2} \int \upd \xb \upd \yb \frac{\hat{M}_{r_c}(\xb)\hat{M}_{r_c}(\yb)}{|\xb-\yb|}.
\end{equation}
This pair potential is exactly the quantum pair potential one would have obtained at the non-relativistic limit of a fully quantum theory of gravity, up to the smearing of the interaction for distances shorter than $r_c$. The additional decoherence terms in \eqref{eq:master4} precisely insure that the resulting interaction remains classical. They would have been non-trivial to guess starting directly from a master equation. This decomposition of semiclassical gravity into 3 decoupled terms --a purely quantum potential, ``intrinsic'' measurement decoherence, and ``gravitational'' feedback decoherence -- is a major advantage of the continuum. The master equation \eqref{eq:master2} we obtained for the discrete model where flashes source the gravitational field, while simpler to derive, had no such simple decomposition.

\subsection{Minimum decoherence prescription}
We constructed a class of non-relativistic theories of semiclassical gravity parameterized by a kernel $\Gamma(\xb,\yb)$. Should we take the CSL kernel which looks simpler, as it is local, or the DP kernel, which is non-local but requires one parameter less? Should we pick yet anoter kernel? Following \cite{tilloy2017}, we may want to find a principle allowing us to single out one kernel. A natural principle is to ask that the total decoherence induced by the model be minimal. Decoherence coming from the measurement and feedback parts can be grouped in a single kernel $\Gamma^\text{f.b.}$
\begin{align}
    \frac{\upd}{\upd t} \rhoav &= -\frac{i}{\hbar}\left[ \hat{V} ,\rhoav \right] -\frac{1}{8}\int  \upd \xb \upd \yb \;\Gamma^\text{f.b.}(\xb,\yb)  \left[\hat{M}_{r_c}(\xb)\left[\hat{M}_{r_c}(\yb),\rhoav\right]\right]
\end{align}
where
\begin{equation}\label{eq:gammafb}
    \Gamma^\text{f.b.}=\, \Gamma + 4 \, (4\pi G/\hbar)^2\; \frac{1}{\nabla^2} \, \Gamma^{-1} \, \frac{1}{\nabla^2}.
\end{equation}
There are many ways to define the minimum for such a kernel. Assuming a translation invariant kernel $\Gamma(\xb,\yb) = \Gamma(\xb-\yb)$ we can go to Fourier space where all terms in \eqref{eq:gammafb} are diagonal. A natural prescription is then to minimize decoherence Fourier mode by Fourier mode, or eingenvalue per eigenvalue. This will insure  that the kernel we find is indeed the minimum for most commonly used norms. Going back to real space, this yields a minimal total decoherence kernel $\Gamma^\text{f.b.}_\text{min}$:
\begin{equation}
    \Gamma^\text{f.b.}_\text{min} (\xb,\yb)= 2 \, \Gamma_\text{min}(\xb,\yb)= \frac{4 G}{\hbar}\frac{1}{|\xb-\yb|}.
\end{equation}
This is precisely the decoherence kernel of the Di\'osi-Penrose model (up to a factor $2$)\footnote{The present approach gives more than the initial Di\'osi-Penrose model. First the derivation of the kernel is not heuristic but motivated by a precise principle: the fact that we get a kernel $\propto 1/|\xb|$ is not by analogy with gravity, it comes naturally. Second, our model does contain an effective gravitational interaction $\hat{V}$, which is not in the historical DP model.}. At the minimum of decoherence, exactly as much decoherence is provided by the continuous measurement and feedback parts.

\section{Consequences and limitations of ``measurement and feedback'' inspired semiclassical models}

\subsection{Non-linear stochastic description and the measurement problem}

An important byproduct of the semiclassical approach we have considered is that it is intimately tied to a solution of the measurement problem. Starting from the will to implement gravity in a semiclassical manner, one is lead to the need to have a quantity formally behaving like a measurement result. Extracting such a result cannot be done without a backreaction on the quantum state, acting in kicks or continuously, that kills off macroscopic superpositions. Requiring that a force is implemented semiclassically, formally as a measurement + feedback scheme naturally leads to objective collapse models. Conversly, taking a collapse model for granted, one gets a way to implement consistent semiclassical interactions ``for free''.

We have focused on feedback master equations averaged over measurement outcomes (or collapse events), which are indeed sufficient to get all possible predictions in all possible experimental situations. But the stochastic description with pure states is important: it is necessary to understand why this approach solves the measurement problem by collapsing macroscopic superpositions. Further, without the initial stochastic state description, there is no way to implement the feedback in the first place, and the empirical content \emph{does depend} on the measurement stochastic Schr\"odinger equation one starts with. Different unravelings of the measurement master equation yield different feedback master equations.

Is it generic that a fundamentally semiclassical theory of gravity is tied to a solution of the measurement problem? I think so. In any reasonable semiclassical theory of gravity, there is a classical configuration associated to spacetime. This classical configuration can be directly given the status of local beable or used, via Einstein's equation, to define a configuration for matter beables. In any case, one has something tangible which can, for example, yield definite measurement outcomes upon coarse graining.

\subsection{Regularization and complete falsifiability}
An undoubtedly unappealing feature of the present approach is the need for an arbitrary spacetime regularization of the mass-density measurement. Sending the smearing length-scale $r_c$ to zero yields a divergent model even for the kernel $\Gamma_\text{min}^\text{f.b.}$ yielding minimal decoherence (Di{\'o}si-Penrose kernel). For standard collapse models, this regularization freedom in principle allows to escape falsifiability indefinitely. In practice, if regularization length-scales get too large, collapse models no longer achieve their purpose of collapsing macroscopic superpositions. This gives a rather soft metaphysical bound that one can easily move by orders of magnitude (see e.g. \cite{feldmann2012}). 

In contrast here, the gravitational potential $\hat{V}$ is inevitably smeared at the same lengthscale as decoherence. Making the regularization lengthscale too large destroys the Newtonian pair potential at short distances. This means that $r_c$ can be experimentally lower bounded by probing intrinsic decoherence and upper bounded by probing gravity at short distances. The semiclassical models we presented are thus falsifiable for all values of $r_c$, although they are so far \emph{not} falsified.

\subsection{Difficult relativistic extension}

The main limitation of the approach I have discussed here is that the construction is made directly in the Newtonian limit. There is so far no complete relativistic theory from which one obtains the Newtonian limit after a robust derivation. What goes in the way of such a construction?

First, constructing a fully relativistic theory of continuous quantum measurement seems difficult. Intuitively, it seems necessary to continuously measure local operators to preserve Lorentz invariance, as was \eg proposed in \cite{diosi1990relat}. The difficulty is that such theories are divergent in a way that does not seem easily curable except with an explicit smearing like the one we used for non-relativistic models. In the Lorentzian context, a space smearing necessarily goes hand in hand with a time smearing, i.e. some form of non-Markovianity. While the theory of non-Markovian stochastic Schr{\"o}dinger equations is well developed \cite{strunz1996,diosi1997,gambetta2002,diosi2014,tilloy2017unrav}, the latter have no real-time continuous measurement interpretation \cite{gambetta2003,diosi2008,diosi2008erratum,wiseman2008} and cannot be used to implement a consistent semiclassical interaction (at least not in any obvious way). There have nonetheless been recent attempts at making general relativistic collapse models with discrete jumps \cite{juarezaubry2017} that may be usable to extend the first toy model we presented. 

Once a proper continuous measurement model or collapse model is available in the general relativistic setting, there is still a difficulty. Generically, the signal obtained will not verify the conservation laws the corresponding field would obey in the classical theory. For example, the ``signal'' stress-energy tensor will not be conserved, which makes it unusable as source in Einstein's equations. This lack of conservation can be dealt with by going to unimodular gravity \cite{josset2017}, but the precise consequences in a complete model are yet to be explored.

A promising avenue of research, in the foreseeable future, is cosmology, where having a complete relativistic model is not necessary. The cosmological consequences of a semiclassical coupling in the spirit of the one we presented here, but for the scale factor, have been explored already by Altamirano \textit{et al.} in \cite{altamirano2017,altamirano2018}.

\section{Conclusion}

The core of the argument in favor of merely considering the option that gravity could be fundamentally semiclassical is that one can construct semiclassical models that are not obviously falsified and that bypass the main no-go argument against semiclassical theories. Admittedly, these models are not  the most appealing on the market, one might even say that they are ugly. Further, they have been made precise only in Newtonian settings: while this is the setting in which criticisms of semiclassical gravity have been formulated, and thus where it is fair to refute them, this is a strong limitation. As a result, the semiclassical models which have been explored in the last 5 years and which I presented here should not be taken too seriously as an alternative to more developed programs like String Theory or Asymptotically safe gravity (to name only a few), but rather as counter examples, as toys, simply reminding us to remain open minded. It is unlikely that semiclassical gravity can be eliminated exploiting purely theoretical arguments, and a hint about the nature of gravity will most likely have to be found in tabletop experiments \cite{bose2017,carney2019}.

\ack{
I thank the participants of the DICE 2018 meeting for interesting and enjoyable discussions. I am  indebted to Lajos Di{\'o}si with whom I have worked on many of the ideas presented here. This work was supported by a postdoctoral fellowship of the Alexander von Humboldt foundation.
}
\section*{References}
\bibliographystyle{iopart-num}
\bibliography{main}

\end{document}